\documentclass[preprintnumbers,amsmath,amssymb]{revtex4}

\usepackage{amsmath}    % need for subequations
\usepackage{graphicx}   % for figures
\usepackage{color}

\pdfoutput=1
\begin{document}

\title{Hartman effect, time-delays, and the non-spatial nature of quantum particles}
\author{Massimiliano Sassoli de Bianchi}
\affiliation{Center Leo Apostel for Interdisciplinary Studies, Vrije Universiteit Brussel, 1050 Brussels, Belgium}\date{\today}
\email{msassoli@vub.be}   %optional

\begin{abstract}

\noindent The phenomenon of quantum tunneling challenges classical intuitions about particle behavior and offers profound insights into the non-spatial nature of quantum entities. This article examines tunneling through the lens of the effect predicted by Thomas E. Hartman in 1962, that the tunneling time-delay becomes independent of the barrier width as the latter increases, implying superluminal effective velocities that appear to conflict with relativistic constraints. Exploiting the inherent symmetries of transmitted and reflected time-delays, our analysis shows that a tunneling particle, like its reflected counterpart, completely avoids the barrier region. However, since it can be detected on the transmitted side of the barrier, the only logical conclusion is that the classical notion of spatiality is violated during the tunneling process. In other words, quantum tunneling and the associated Hartman effect strongly support the idea that quantum non-locality should be interpreted as quantum non-spatiality. Also, according to the conceptuality interpretation of quantum mechanics, proposed by Diederik Aerts in 2009, a non-spatial state can be viewed as an abstract state of a conceptual-like entity, which only enters the spatiotemporal domain of our physical reality when it reaches its maximum degree of concreteness, during the wave-function collapse. In this framework, a scattering process becomes analogous to a cognitive-like interrogative process, where the available answers are the transmitted and reflected outcomes. Since the time required to answer a question is generally not directly related to the processes described in the possible answers, this provides insight into why the quantum time-delays of transmission and reflection can be similar or even identical, although from a classical spatial perspective they should be meaningfully different.
\end{abstract} 

\maketitle
\\

\section{Introduction}

\noindent \emph{Hartman effect} is the theoretical prediction  that the average time-delay experienced by a quantum entity tunneling through a potential barrier is independent of its width, in the limit where the latter becomes increasingly larger \citep{Hartman1962}.  Numerous experiments have been carried out over the years, the more recents exploiting the progresses made in ultrafast lasers and attosecond metrology \citep{Hentschel2001}.  Essentially, they have  confirmed the reality of Hartman effect, i.e., the possibility of effective superluminal velocities \citep{Enders1992,Enders1993,Steinberg1993,Spielmann1994,Longhi2001,Landsman 2014, Torlina2015,Camus2017,Sainadh2019,Wang2000,Yu2022}.  There have also been some criticisms about what some of these experiments would have actually shown  \citep{Winful2006,Sokolovski2018}, considering that the interpretation of the results is not straightforward and that the tunneling process is complex, sensitive to the specifics of the potential barrier, the nature of particles interacting with it and the methodology each time used. 

Although it is good to be cautious, considering the volume of experimental results, it also gets harder and harder to deny that what the theory predicts has been essentially validated: that tunneling is an almost instantaneous  process. It would certainly not be the first time that a quantum phenomenon that is difficult to interpret would be experimentally confirmed and then accepted, despite initial resistance. Quantum physics, since its inception, has deeply shaken our classical prejudices, validating phenomena such as superposition \citep{Rauch1975,Gerlich2013}, entanglement \citep{christensen2013,hensen-etal2016} and indistinguishability \citep{Bradley1995,Davis1995}. 

The position that we will adopt in this article is that Hartman effect has been essentially verified, and that future experiments will close the loopholes of the current ones, confirming the predictions of the quantum formalism. Hence, it is important to try to understand what these results tell us about the nature of the microphysical world. This is what we will do in the following, emphasizing that the tunneling phenomenon urges us to embrace the concept of \emph{quantum non-spatiality} as a fundamental aspect of quantum nature and behavior. Of course, tunneling and the associated Hartman effect is not the only quantum phenomenon  that requires us to embrace such a paradigm shift, but surely it does so in a very direct and convincing way, possibly changing the minds of those less inclined to abandon the classical prejudice that our physical reality would be solely of a spatiotemporal nature \citep{Aerts1999,Sassoli2021}. 

The article is organized as follows. In Sec.~\ref{Superluminal}, we recall the usual statement of Hartman effect and explain why it is associated with an infinite effective velocity. In Sec.~\ref{conditional}, we emphasize that although there is no uncontroversial notion of \emph{conditional sojourn time} in quantum mechanics, one can still make sense of a notion of \emph{conditional time-delay}. In Sec.~\ref{three-step}, we analyze the reflection of a particle by a potential barrier, in conjunction with the semiclassical approximation and the unitarity of the scattering matrix, to gain insight into the quantum tunneling phenomenon, showing that the time-delay experienced by a tunneling entity is identical to that of a reflected one. In Sec.~\ref{Solving}, we describe our solution to the conundrum posed by Hartman effect, suggesting that a tunneling entity has to be understood as a non-spatial entity, only actualizing its presence in space when ultimately detected. In Sec.~\ref{conceptuality}, we briefly introduce Aerts' \emph{conceptuality Interpretation} of quantum mechanics, as a possible explanation of the nature of a non-spatial entity. Finally, in Sec.~\ref{Conclusions}, we summarize the essential points of what has been presented.

\section{Superluminal effective velocity}
\label{Superluminal}

\noindent Limiting our discussion to a single dimension of space, we consider an entity impinging on a potential barrier, moving from left to right. When it crosses the barrier, due to the quantum tunneling effect, its motion becomes apparently superluminal, in the sense that to account for the very large time-advance (i.e., negative time-delay) it accumulates with respect to a free reference entity with same incoming state, it becomes necessary to attribute to the interacting entity an effective velocity within the barrier that tends to infinity, as the barrier width increases. 

\begin{figure}[!ht]
\centering
\includegraphics[scale =0.38]{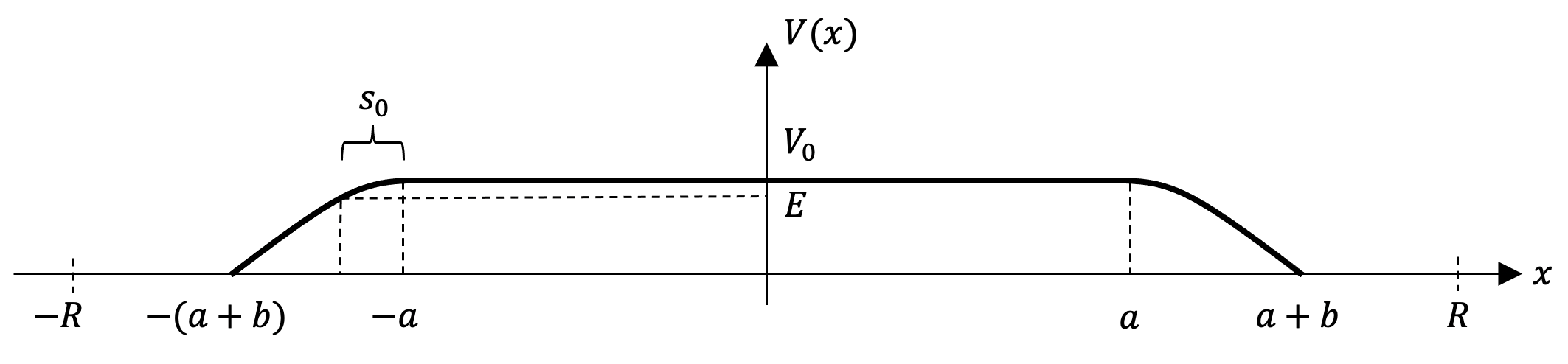}
\caption{A potential barrier whose region of constant height $V_0$ has width $2a$, while the left and right transition regions have each width $b$, hence $V(x)$ has its support in the interval $[-a-b,a+b]$, contained in a larger interval $[-R,R]$, $R>a+b$, which is the one considered for the calculation of the time-delay. If the energy $E={1\over 2}mv^2$ of a classical particle coming from the left is strictly below the barrier height, $E<V_0$, it can only reach point $x=-a-s_0$, with $V(-a-s_0)-E=0$, before being reflected back.  
\label{figure1}}
\end{figure}

To see this, let $2a$ be the width of the potential barrier corresponding to the region where it has constant height $V_0$, with $2(a+b)$ the length of the overall interval where it has its support, including the two transition regions (see Fig.~\ref{figure1}). Then, Hartman effect is typically expressed as the following asymptotic behavior of the \emph{transmission time-delay} $\tau_{\rm tr}$, as the barrier width $a$ increases: 
\begin{equation}
\tau_{\rm tr}=-{2a\over v}\left[1 + {\rm O}\left({1\over a}\right)\right]
\label{Hartman}
\end{equation}
where $v$ is the incoming velocity, here assumed to be positive, since the entity comes from the left. 

By definition of time-delay (see also the discussion in Sec.~\ref{conditional}), and reasoning as if we were dealing with an entity of a corpuscular nature (it is obviously not correct to do so, but at the moment this will help us in the reasoning), the time $T_{\rm tr}$ the particle spends within the barrier region is given by the time-delay (\ref{Hartman}) plus the time a free reference particle of same incoming velocity spends in that same region, i.e., 
\begin{equation}
T_{\rm tr}=\tau_{\rm tr}+{2(a+b)\over v}
\label{time-of-sojourn}
\end{equation}
Inserting (\ref{Hartman}) into (\ref{time-of-sojourn}), we find that, as $a\to\infty$, $T_{\rm tr}$ tends to a constant, which means that the effective average velocity $v_{\rm eff}$ inside the barrier, 
\begin{equation}
v_{\rm eff}={2(a+b)\over T_{\rm tr}}
\label{edffective-velocity}
\end{equation}
tends to infinity, as $a\to\infty$.

The probability for a classical corpuscle of mass $m$ and velocity $v=\sqrt{2E/m}$, with $E<V_0$, to be transmitted through the barrier, is of course zero, hence interpreting the tunneling phenomenon using a classical analogy is not sufficient, since only quantum entities can be detected on the right hand side of the barrier with non-zero probability, when the incoming energy $E$ is strictly below the barrier height $V_0$. But even if we renounce viewing the tunneling entity as a corpuscle, the question of the effective velocity of the process remains. How can a quantum entity, impinging on the barrier from the left, reappear almost instantly to its right, so that its effective velocity $v_{\rm eff}$ becomes arbitrarily large? 

It is important to observe that superluminal phenomena exist and pose no problems in physics. A famous example is the speed of the shadow of an object \citep{Leblond2006}, which is not limited by any physical principle and therefore has no upper limit. Tunneling, as a superluminal phenomenon, is like a shadow: it cannot be used to produce faster-than-light signaling, considering in particular that the tunneling probability is very small; see for instance the discussion in \citep{Dumont2020}. That said, to what extent can we say that the tunneling process corresponds to a propagation through space at an effective speed greater than the speed of light?

Let us immediately mention the possible criticism that Hartman effect is typically derived in the ambit of non-relativistic quantum mechanics, whereas in relativity speeds are limited by the light speed $c$. Many studies have already addressed such criticism, showing that superluminal effective tunneling velocities, akin to Hartman effect, also appear when using relativistic equations, like the Dirac equation \citep{DeLeo2007,DeLeo2013}.  Another typical criticism is that there would be no uncontroversial notion of tunneling time \citep{Buttiker1982,Buttiker1983,Hauge1989,Landauer1994,Carvalho 2002,Olkhovksy 2004,Muga2008,Sassoli2012,Sokolovski2018,field2022}, hence one could not meaningfully talk of the tunneling entity's speed in the barrier region. We address this second criticism in some detail the next section, which however is not essential to understanding the content of this article, and can be safely skipped on a first reading.

\section{Conditional time-delay}
\label{conditional}

\noindent It is well known that there is no intrinsic arrival time operator in quantum mechanics \citep{Allcock1969,Muga2008}. However, the notion of \emph{sojourn time} (also called \emph{dwell time}), defined in terms of probabilities of presence, remains valid also in quantum mechanics, i.e., can be associated with a bona fide self-adjoint operator \citep{Martin1981,Jaworski1989}. But to associate a time-delay to a tunneling entity, one needs a more specific notion of \emph{conditional time-delay}, which in turn requires a notion of \emph{conditional sojourn time}, that is, of time of sojourn conditional to the fact that the entity is ultimately observed in a given state, in our case, a transmitted or reflected state. 

Unfortunately, one cannot define a quantum conditional sojourn time operator, providing a meaningful answer to the question of how much time, on average, an entity has passed inside a given region of space, say of radius $R$, incorporating the potential barrier (see Fig.~\ref{figure1}), conditional to the fact that it is ultimately observed in a given asymptotic state (transmitted or reflected). This because it requires to jointly answer two incompatible questions: one about the presence of the entity inside said region, and one about knowing if it will be ultimately transmitted or reflected. Let us show this in some detail. The question about the spatial localization is associated with the projection operator (here in Dirac notation): 
\begin{equation}
P_R=\int_{-R}^R dx |x\rangle\langle x|
\end{equation}
If $|\psi_t\rangle=e^{-{i\over\hbar}Ht}\Omega_-|\phi\rangle$ is the scattering state describing the incoming entity at time $t$,  where $H=H_0+V$ is the total Hamiltonian, $H_0={P^2\over 2m}$ the free Hamiltonian, $\Omega_\pm=\lim_{t\to\infty}e^{\pm{i\over\hbar}Ht}e^{\mp{i\over\hbar}H_0t}$ the wave operators and $|\phi\rangle$ the incoming state at time $t=0$, then the average $\langle \psi_t |P_R|\psi_t\rangle$ is the probability of finding the entity in question, at time $t$, inside $[-R,R]$. Integrating such probability of presence over all times, one obtains the so-called \emph{global} (unconditional) \emph{sojourn time} in the region $[-R,R]$: 
\begin{equation}
T_R(\phi)=\int_{-\infty}^\infty dt \langle \psi_t |P_R|\psi_t\rangle
\end{equation}
The \emph{global time-delay} $\tau(\phi)$, for the initial state $|\phi\rangle$, is then defined as the limit of the difference between the interacting and free global sojourn times, as the localization region covers the entire space \citep{Sassoli2012}: 
\begin{equation}
\label{global}
\tau(\phi)=\lim_{R\to\infty} \left[T_R(\phi)-T_R^0(\phi)\right]
\end{equation}
where the free reference time is given by $T_R^0(\phi)=\int_{-\infty}^\infty dt \langle \phi_t |P_R| \phi_t\rangle$, with $|\phi_t \rangle =e^{-{i\over\hbar}H_0t}|\phi\rangle$.

On the other hand, the question about the asymptotic behavior is formally associated with the projection operator $F_t^{\rm tr}$, for transmission, and $F_t^{\rm re}$, for reflection, given by: 
\begin{equation}
F_t^{\rm tr}=e^{-{i\over\hbar}Ht}\Omega_-S^\dagger F^{+}S\Omega_-^\dagger e^{{i\over\hbar}Ht},\quad F_t^{\rm re}=e^{-{i\over\hbar}Ht}\Omega_-S^\dagger F^{-}S\Omega_-^\dagger e^{{i\over\hbar}Ht}
\end{equation}
where $F^{+}=\Theta(+ P)$ and $F^{-}=\Theta(- P)$ are the projection operators onto the set of states of positive and negative momentum, respectively, and $S=\Omega_+^\dagger\Omega_-$ is the unitary scattering operator. It is not important whether the time-dependent observables $F_t^{\rm tr}$ and $F_t^{\rm re}$ can be measured in practice, what matters here is that they can be consistently defined and measured in principle, and that the averages $P_{\rm tr}(\phi)=\langle \psi_t |F_t^{\rm tr}|\psi_t\rangle=\langle \phi |S^\dagger F^{+} S|\phi\rangle$ and $P_{\rm re}(\phi)=\langle \psi_t |F_t^{\rm re}|\psi_t\rangle=\langle \phi |S^\dagger F^{-} S|\phi\rangle$ correctly give the transmission and reflection probabilities for the incoming state $| \phi\rangle$, here assumed coming from the left, i.e., $F^{+}|\phi\rangle=|\phi\rangle$. 

So, for reflection (respectively, transmission), we have two projection operators, $P_R$ and $F_t^{\rm re}$ (respectively, $F_t^{\rm tr}$), that allow to test, at time $t$, the following two properties: ``the quantum entity is in region $[-R,R]$'' and ``the quantum entity is reflected (respectively, transmitted) by the potential barrier.'' However, since $[P_R, F_t^{\rm re}]\neq 0$, one cannot define a projection operator testing, at time $t$, the meet property ``the quantum entity is in the region $[-R,R]$ \emph{and} is reflected by the potential barrier,'' and the same holds for the transmission case. One can however observe that since $P_R\to\mathbb{I}$, as $R\to\infty$, one has $[P_R, F_t^{\rm re}]\to 0$, as $R\to\infty$, i.e., the two observables $P_R$ and $F_t^{\rm re}$ become compatible in the time-delay limit. One can thus define the auxiliary symmetric operator $W_{R,t}^{\rm re}={1\over 2}(P_R F_t^{\rm re}+F_t^{\rm re}P_R)$ and observe that the average $\langle \psi_t |W_{R,t}^{\rm re}|\psi_t\rangle$ has all the good properties of a joint probability, except that it can also take negative values. However, as $R\to\infty$,  it becomes non-negative and one can still define a bona fide notion of (conditional) time-delay for the reflected outcome, as the limit of the difference:
\begin{equation}
\label{time-delay-limit-refl}
\tau_{\rm re}(\phi)=\lim_{R\to\infty} \left(T_R^{\rm re}(\phi)-T_R^{0,\rm re}(S\phi)\right)
\end{equation}
where we have defined the auxiliary interacting and free sojourn times associated with the reflection outcome: 
\begin{equation}
\label{cond-sojourn}
T_R^{\rm re}(\phi)=P^{-1}_{\rm re}(\phi)\int_{-\infty}^\infty dt \langle \psi_t |W_{R,t}^{\rm re}|\psi_t\rangle,\quad T_R^{0,\rm re}(S\phi)=P^{-1}_{\rm re}(\phi)\int_{-\infty}^\infty dt \langle S\phi_t |W_{R,t}^{0,\rm re}| S\phi_t\rangle
\end{equation}
where $W_{R,t}^{0,\rm re}={1\over 2}(P_R F^{-}+F^{-}P_R)$. The transmission time-delay $\tau_{\rm tr}(\phi)$ is similarly defined, replacing in the different expressions $F^{-}$ by $F^{+}$. Note that the free reference time is now defined in terms of the outgoing free evolving state $| S\phi_t\rangle=e^{-{i\over\hbar}H_0t}S|\phi\rangle$, which is necessary in order to correctly subtract the contribution in $T_R^{\rm re}(\phi)$ that linearly diverges, as $R\to\infty$. Studying the limit (\ref{time-delay-limit-refl}) is beyond the scope of this article and we refer for this to \citep{Sassoli1993,Sassoli2012}. For well behaved potential functions and incoming states, one finds:
\begin{equation}
\label{cond-sojourn2}
\tau_{\rm re}(\phi)=P^{-1}_{\rm re}(\phi)\int_{0}^\infty dE  |{\cal L}(E) \phi(E)|^2  \hbar {d\alpha_{\cal L}(E)\over dE}
\end{equation}
where ${\cal L}(E)=|{\cal L}(E)|e^{i\alpha_{\cal L}(E)}$ is the reflection amplitude from the left at energy $E$. Considering the mono-energetic limit $|\phi(E')|^2\to\delta(E'-E)$, one then finds that the reflection time-delay for an entity coming from the left, with energy $E$, is given by: 
\begin{equation}
\label{cond-sojourn3}
\tau_{\rm re}^{\rm left}(E)=\hbar {d\alpha_{\cal L}(E)\over dE},
\end{equation}
and similar formulae can be written for the reflection from the right and for the transmission time-delays; see Eq. (\ref{3-time-delays}).

Concluding this brief analysis, which is not meant to be exhaustive, one can still make sense of a notion of transmission and reflection time-delays \citep{Sassoli1993,Sassoli2012,Jaworski1988}, and more generally of angular time-delays \citep{Bolle1976}, compatibly with those approaches that use a direct analysis of the evolution of wave packets \citep{Hauge1989,Jaworski1988,Sassoli2012}. Having clarified this point, we want now to use the notions of quantum transmission and reflection time-delays in the analysis of the tunneling phenomenon, to extract information about the nature of a quantum entity.

\section{A three-step reasoning}
\label{three-step}

The reasoning we will develop is in three steps. Step 1 is a careful analysis of the process of a classical entity which is reflected from the barrier. Step 2 is the use of the semiclassical approximation. Step 3 is the exploitation of a remarkable symmetry property of the quantum $S$-matrix, which allows us to deduce that what happens for the reflection from the barrier also holds for the transmission (tunneling) ``through'' it. 

Our starting point is the calculation of the reflection time-delay experienced  by a classical corpuscle interacting with the potential barrier of Fig.~\ref{figure1}: 
\begin{equation}
\label{potential}
V(x)=V_0
\begin{cases}
1 & |x| < a\\
h(|x|-a) & a\leq |x|\leq a+b
\end{cases}
\end{equation}
where $a>0$ and $h(s)$ is a smooth monotone decreasing function with support in the interval $[0, b]$, $b>0$, with $h(0) = 1$ and $h(b)=0$, describing the switching on and off of the barrier in space. A classical particle of energy $E<V_0$ is always reflected back from the potential barrier. According to (\ref{global})-(\ref{time-delay-limit-refl}), to calculate the associated time-delay, one needs to calculate the difference between the time $T_{\rm re}^{\rm cl}$ it spends inside a region of radius $R>a+b$, and the time $T^{\rm cl}_0={2R\over v}$ spent in that same region by a free reference particle of same energy $E={1\over 2}mv^2$, then take the limit $R\to\infty$, which will be trivial here, the potential being compactly supported: 
\begin{equation}
\tau_{\rm re}^{\rm cl}=\lim_{R\to\infty}(T_{\rm re}^{\rm cl} -T^{\rm cl}_0)
\label{time-delay-classical}
\end{equation}
More precisely, a simple calculation gives for  the classical reflection time-delay \citep{Narnhofer1981,Sassoli2000}:
\begin{equation}
\tau_{\rm re}^{\rm cl}=\left[2\int_{-b}^{-s_0}{ds\over v(s)}-{2b\over v}\right]+\left[0-{2a\over v}\right]
\label{time-classical}
\end{equation}
where 
\begin{equation}
v(s)=\sqrt{2[E-V_0h(-s)]\over m}
\label{time-classical2}
\end{equation}
and $s_0$ is such that $E-V_0h(s_0)=0$, so that $x=-a-s_0$ is the extreme point reached by the particle inside the potential region, before being reflected back; see Fig.~\ref{figure1}. 

To understand the physical content of (\ref{time-classical})-(\ref{time-classical2}), we must note that the time spent by the free reference particle inside the potential region is ${2(a+b)\over v}$. The term ${2a\over v}$ describes the time spent in the region where the potential is constant and the term ${2b\over v}$ the time spent in the two transition regions, where there is a non-zero gradient. One can also think of the free reference particle as a particle that is elastically reflected at the origin, but in both cases it will have to cross the variable gradient transition region twice: once going in, and once coming out. Now, since the incoming energy $E$ is below the maximum height $V_0$ of the potential barrier, the particle cannot explore its region of length $2a$ where it is constant, hence, the time it spends inside that region is exactly zero, which is the reason of the $0$-term indicated in the second bracket in (\ref{time-classical}), which quantifies the time-delay produced by the zero-gradient part of the potential. 

On the other hand, the first bracket in (\ref{time-classical}) is the contribution to the time-delay due to the transition region, which is explored twice by the reflected particle, first when it moves from the left to the right, until it reaches the extreme point of its movement at $x=-a-s_0$, and then when it comes back from where it came, having reversed its motion. Clearly, the expression in the first bracket does not depend on the barrier width $2a$, but only on the detail of the transition region, hence we have the following asymptotic form, for large values of $a$: 
\begin{equation}
\tau_{\rm re}^{\rm cl}=-{2a\over v}\left[1 + {\rm O}\left({1\over a}\right)\right]
\label{Hartman-ref}
\end{equation}

We can observe that (\ref{Hartman-ref}) is identical to (\ref{Hartman}). However, different from (\ref{Hartman}), there is no mystery here in the observed asymptotic behavior. It comes from the second bracket in (\ref{time-classical}), which is a consequence of the fact that the reflected particle never enters the (constant height) barrier region of width $2a$. Note also that it would be meaningless to use (\ref{Hartman-ref}) to speak of an effective velocity of the reflected particle inside the barrier region of width $2a$, as is clear that it never enters such region. 

What is the relevance of the above to the problem of interpreting Hartman effect in quantum tunneling? As we said, two additional steps will allow us to make the connection: the fact that classical time-delays correspond to the semiclassical approximations of their quantum mechanical analogues \citep{Narnhofer1981,Berry1982,Fedoriouk1987,MartinSassoli1994}, and the existence of a remarkable relation between the transmission and reflection time-delays in quantum mechanics, which follows from the unitarity of the \emph{scattering matrix} $S(E)$ \citep{SassoliDiVentra1995}. 

Let us start by observing how unitarity affects the quantum time-delays. For a one-dimensional problem, since there are only two directions, the energy shell scattering matrix for energy $E$ is the $2\times 2$ matrix:
\begin{equation}
S(E)=
\left[ \begin{array}{cc}
{\cal T}(E) & {\cal R}(E) \\
{\cal L} (E)& {\cal T}(E) \end{array} \right]
\label{ScatteringMatrix}
\end{equation}
where ${\cal T}(E)$ is the transmission amplitude and ${\cal L}(E)$ and ${\cal R}(E)$ are the reflection amplitudes from the left and right, respectively, for energy $E$. It is then straightforward to observe that the unitarity relation, $S^\dagger (E)S(E) = \mathbb{I}$, implies probability conservation
\begin{equation}
|{\cal T}(E)|^2 +|{\cal R}(E)|^2 = |{\cal T}(E)|^2 + |{\cal L}(E)|^2 = 1
\end{equation}
as well as the relation 
\begin{equation}
{\cal T}^*(E){\cal R}(E) + {\cal L}^*(E){\cal T}(E)=0
\end{equation}
which can be equivalently written, using $|{\cal R}(E)| = |{\cal L}(E)|$, as the phase relation:
\begin{equation}
\alpha_{\cal T}(E)+ {\pi\over 2} ={1\over 2}\left[\alpha_{\cal L}(E) +\alpha_{\cal R}(E)\right] \mod\pi.
\label{phase}
\end{equation}
where $\alpha_{\cal T}(E)=\arg {\cal T}(E)$, $\alpha_{\cal L}(E)=\arg {\cal L}(E)$ and $\alpha_{\cal R}(E)=\arg {\cal R}(E)$.

Following (\ref{cond-sojourn3}), the energy shell quantum mechanical transmission and reflection time-delays, for energy $E$, are given by the reduced Planck constant $\hbar$ times the energy derivatives of the phases of the transmission and reflection amplitudes \citep{Eisenbud1948,Wigner1955,Smith1960,Sassoli2012}:
\begin{equation}
\label{3-time-delays}
\tau_{\rm tr}(E)=\hbar {d\alpha_{\cal T}(E)\over dE}\quad\quad \tau_{\rm re}^{\rm left}(E)=\hbar {d\alpha_{\cal L}(E)\over dE}\quad\quad \tau_{\rm re}^{\rm right}(E)=\hbar {d\alpha_{\cal R}(E)\over dE}
\end{equation}
and from (\ref{phase}) we obtain the remarkable relation: 
\begin{equation}
\tau_{\rm tr} (E)={1\over 2}\left(\tau_{\rm re}^{\rm left} (E)+\tau_{\rm re}^{\rm right}(E)\right)
\label{time-relation}
\end{equation}
In other words, the transmission time-delay is the arithmetic average of the reflection time-delays from the left and from the right, and since in our simplified analysis we have considered a symmetric potential barrier, we have $\tau_{\rm re}^{\rm left} (E)=\tau_{\rm re}^{\rm right}(E)$ and (\ref{time-relation}) becomes: 
\begin{equation}
\tau_{\rm tr} (E)= \tau_{\rm re}(E)
\label{time-relation2}
\end{equation}

Note that (\ref{time-relation}) and (\ref{time-relation2}) are not approximations, but exact identities, expressing a fundamental difference between quantum mechanics and classical mechanics. Indeed, in classical mechanics, for a given non-zero incoming energy $E$, a particle is always either transmitted or reflected, so relations of the above kind would be meaningless for a spatial corpuscle. On the other hand, in quantum mechanics both outcomes, transmission and reflection, are always possible, hence transmission and reflection time-delays can be jointly defined for a same incoming energy $E$, and the unitarity of the evolution process then forces them to be equal, in the sense of (\ref{time-relation}) and (\ref{time-relation2}).

Now comes the last step of our reasoning. As we mentioned already, the classical expression (\ref{Hartman-ref}) also holds for a quantum process in the semiclassical regime, i.e., when the de Broglie wavelength $\lambda$ of the incoming entity is sufficiently small compared to the size $b$ of the potential gradient in the transition regions, i.e., if $\lambda \ll  b$, and this all the more so if the function $h(s)$ describing the transition regions is highly differentiable \citep{Narnhofer1981,Berry1982,Fedoriouk1987,MartinSassoli1994}. 

It then follows that (\ref{time-classical})-(\ref{Hartman-ref}) are good approximations for the quantum mechanical transmission time-delays, and this in itself constitutes a derivation of Hartman effect valid for potential barriers of general shape, when the barrier's transition regions vary over distances much larger than the de Broglie wavelength of the incoming entity \citep{Sassoli2000}. However, what interests us is not the derivation in itself, but the insight it allows us to gain into the temporal aspects the tunneling phenomenon. 

\begin{figure}[!ht]
\centering
\includegraphics[scale =0.5]{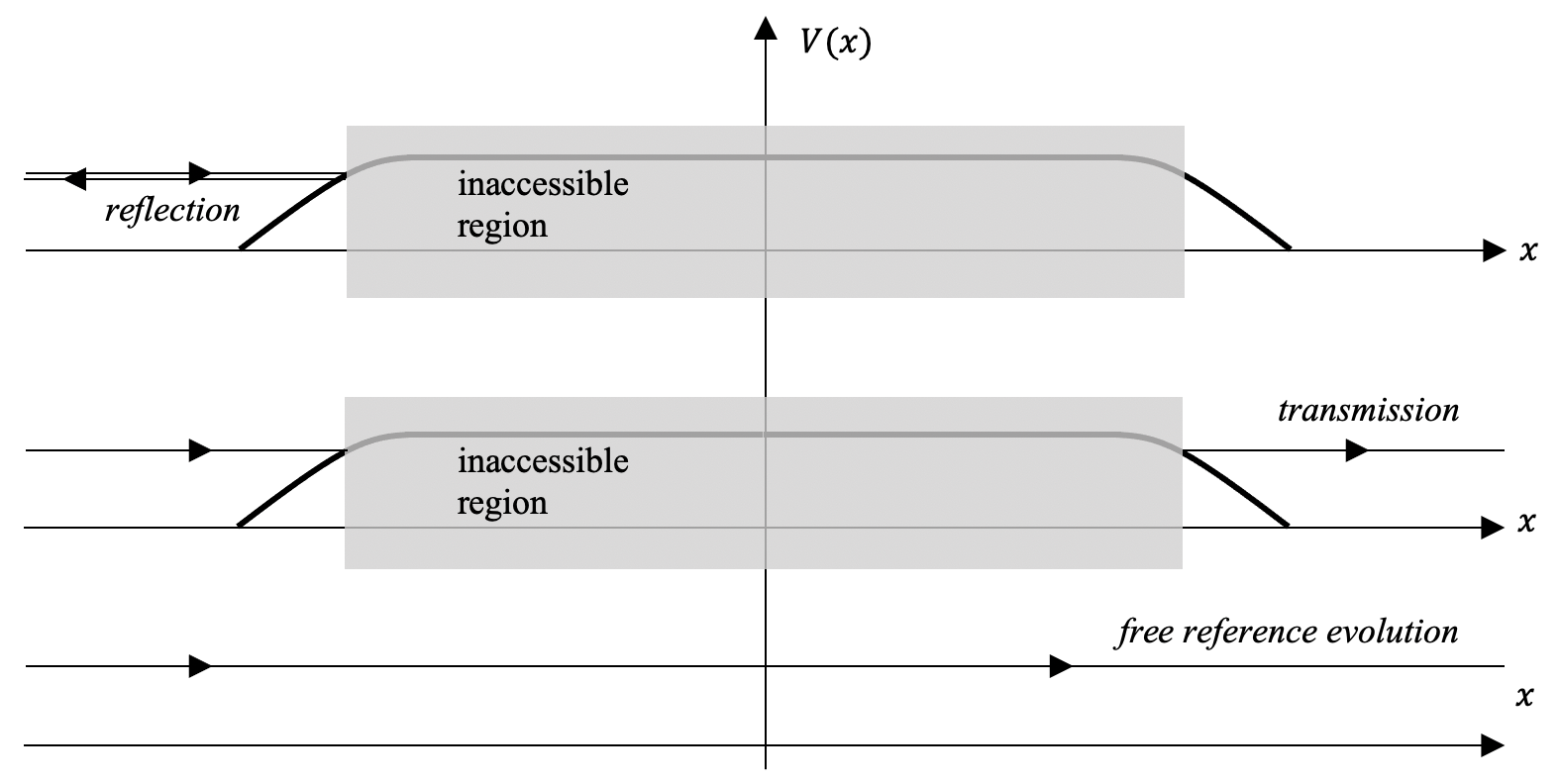}
\caption{Unlike the reference entity that evolves freely, the potential region for which $E-V(x)<0$ remains inaccessible both to the reflected and the transmitted (tunneling) entities. Outside of this inaccessible region, the scattering entities interact for the same amount of time, on average, with the transition regions, whether they are ultimately reflected or transmitted. In the first case, they interact twice with the left transition region (assuming they come from the left), in the second case they interact once with both transition regions, which are here identical being the barrier symmetric. For non-symmetric barriers, the same applies if one considers an average over the processes where the entity approaches the barrier from the left and from the right, respectively, as per (\ref{time-relation}).
\label{figure2}}
\end{figure}

Hartman effect (\ref{Hartman}), when viewed from the perspective of a reflected entity (\ref{Hartman-ref}), is just the consequence of the fact that the latter does not penetrate the entire barrier width, but only interacts with its external transition region. Hence, in its race with a free reference entity, it has to travel a much shorter distance (roughly, shorter than $2a$), which explains its considerable time-advance, i.e., its negative time-delay. From the exact equality (\ref{time-relation2}), it then follows that, mutatis mutandis, the same must also be true for the tunneling entity, i.e., the same mechanism is expected to also explain the value of the transmission time-delay. In other words, the time-advance experienced by the tunneling entity must also be traced back to the fact that, in its race, it  avoids altogether the energetically forbidden region of width $2a$, in exactly the same way it is avoided by the reflected entity. Schematically, we have the situation depicted in Fig.~\ref{figure2}. 

\section{Solving the conundrum}
\label{Solving}

\noindent At this point, we are faced with the following conundrum. Just as it would be absurd to think of the reflected entity as ``passing through'' the potential barrier, the same applies to the tunneling one. The difficulty with the latter statement is that since the tunneling entity approaches the potential barrier from the left, then is detected to its right, and in between there is an inaccessible region that is not crossed (otherwise it would affect the resulting transmission time-delay, making it different from the reflection time-delay), and that there are no other paths in space to go from the left to the right of the potential barrier, one might rightly ask: How a spatial entity, moving in space, can suddenly skip an entire region of arbitrary size, as if it did not exist, and instantly reappear on the other side of it?  

What we want to emphasize is that it is one thing to think of an entity that, for unclear reasons, would be able to move through a potential region with an arbitrary effective velocity, and quite another to observe that, in its evolution, in no way it penetrates, and therefore experiences, such region. In our view, the conundrum has only one solution: we have to think of the tunneling entity not as a spatial entity, permanently present in space, but as a non-spatial entity, only ephemerally present in space. Indeed, a non-spatial entity can actualize its presence in different regions of space, independently of the existence of inaccessible regions separating them. But what exactly do we mean by the term \emph{non-spatiality}? Quoting from \citep{Aerts1999}, non-spatiality is the observation that \citep{Aerts1999}: 
\begin{quote}
Reality is not contained within space. Space is a momentaneous crystallization of a theatre for reality where the motions and interactions of the macroscopic material and energetic entities take place. But other entities -- like quantum entities for example -- `take place' outside space, or -- and this would be another way of saying the same thing -- within a space that is not the three-dimensional Euclidean space.
\end{quote}

\noindent Note that the non-spatiality hypothesis was firstly introduced by Diederik Aerts in the late eighties \citep{Aerts1990} and has since been discussed in a number of works \citep{Aerts1998,Aerts1999}, with other authors also realizing the need for its adoption, to fully explain quantum mechanics \citep{Christiaens2003,Kastner2012,Sassoli2021}. Therefore, the analysis we have here presented only constitutes new evidence that adds to the numerous that have been accumulated over time, telling us that our physical reality would be fundamentally non-spatial  \citep{Sassoli2021}. 

Just to mention two examples, \emph{quantum measurement} and the \emph{Born rule} can be explained by introducing ``hidden'' interactions that are genuinely non-spatial \citep{asdb2014}; also, the phenomenon of quantum entanglement, when understood as correlations created from ``hidden'' connections, requires again the latter to be non-spatial in nature \citep{asdb2016}.

\section{The conceptuality interpretation}
\label{conceptuality}

\noindent In the previous sections, we offered an analysis that illuminates the tunneling phenomenon and associated Hartman effect from a fresh vantage point, asking us to reconsider our preconceptions about the nature of quantum entities, in the sense that their non-locality should be understood as originating from their non-spatiality. An interesting question then imposes itself: What would be the nature of a non-spatial entity, and how does spatiality emerge from a non-spatial reality? Answering this question is important since simply saying that certain processes take place outside of spacetime is obviously not sufficient, if then we can offer no possible explanations of what this might mean. 

This article is not devoted to the analysis of the notion of non-spatiality per se, and more information about it can be found in the articles cited in the previous section. However, it might be useful to also provide an element of a possible answer here. Such element, highly speculative for the time being, but undoubtedly also suggestive, is contained in the \emph{conceptuality interpretation} of quantum mechanics 
\citep{Aerts2009, Aerts2010a,Aerts2010b,Aerts2014,AertsBeltran2020,Aertsetal2020,Aertsetal2024a,Aertsetal2024b}, which emerged from the research program called \emph{quantum cognition}  \citep{aertsaerts1995,aertsbroekaertsmets1999,khrennikov1999,gaboraaerts2002,altmanspacher2002,aertsczachor2004,AertsGabora2005,bb2012}, using the quantum mathematics and associated conceptual structure as a framework to explain human cognition, including judgment and decision making, concepts, reasoning, memory, and perception. 

The unexpected success of the quantum cognition program led one of its pioneers, Diederik Aerts, to cultivate the bold idea that it could not be just a mere coincidence that the mathematical formalism of quantum mechanics was so well equipped in describing so many aspects of the human cognitive domain. The hypothesis that then emerged is that there had to be a deeper correspondence, related to the true nature of the microphysical entities, and more precisely that the latter were to have a nature similar to that of human conceptual entities. 

It is important here to immediately emphasize that we are not equating the entities of the microworld with the concepts of the human cognitive domain. This would be just as wrong as to confuse acoustic waves with electromagnetic waves. Instead, it's about acknowledging that just as there are physical phenomena which, although very different from each other, can share common intrinsic characteristics, explaining the similarities in their behavior, for example a same undulatory nature, the same goes for the conceptual nature, which would be present in the physical world at multiple levels. If this is true, then it becomes possible to understand the non-spatiality of microphysical entities in the same way that we understand the \emph{abstractness} of human concepts, in the sense that a departure from the spatial (and more generally, spatiotemporal) theater would correspond to a change of state equivalent to a growth in abstractness of a conceptual entity. 

In fact, entities of a conceptual nature can find themselves in states of greater or lesser abstractness, or equivalently, of lesser or greater concreteness, and the spatiotemporal layer would be nothing more than a theater inhabited by conceptual entities in a state of maximal concreteness, which leads them to behave similarly to what we call \emph{objects}. In other words, in the conceptuality interpretation a maximally concrete state is considered to describe a situation of maximal localization in space, whereas a maximally abstract state is considered to describe a situation of maximal de-spatialization.

It is obviously not possible to appreciate the explicative power of the above statements without going into the merits of the modeling of human concepts by means of the quantum formalism \citep{asdb2016b} and without describing how more concrete concepts are formed in the human conceptual domain, the matter being also complicated by the fact that, for historical reasons, in the human conceptual domain there are different possible lines connecting \emph{abstract} to \emph{concrete}, the fundamental being the one that expresses the passage from \emph{one-word concepts} to \emph{stories formed by the combination of multiples concepts},  which mirrors the formation of macroscopic objects in the physical world, by aggregation of multiple microscopic entities  \citep{Aertsetal2020}.

What we just wanted to emphasize here is that non-spatiality is not to be understood as an extension of the spatiotemporal theatre to which we should add further dimensions in which the microphysical entities would reside, when they are in superposition states. If the hypothesis of the conceptuality interpretation turns out to be correct, the picture that emerges would be much more complex and articulated, our physical domain being much more fluid and contextual than we had initially imagined, with continuous intertwining and passages between spatiality and non-spatiality, abstractness and concreteness, as it happens in the human conceptual domain, when we describe the interaction between cognitive entities and the entities of meaning forming their languages.

Having said that, and being aware that what follows cannot be fully appreciated without digesting all the nuances of the conceptuality hypothesis (which is easily misunderstood), we would like to briefly explain why the Hartman effect does not present any interpretive difficulties from a conceptuality perspective. If we consider that the (here one-dimensional) scattering process is similar to an interrogative process with two possible answers, we can describe the measuring devices as a whole as a cognitive entity that has to answer a specific question, which in human language we could formulate as follows: ``What is a good example of an outcome for an entity of energy $E$ coming from the left interacting with a barrier of shape $V(x)$?'' Obviously, the relative frequencies of the two possible answers (transmitted or reflected) depend in a complex way on how an entity of energy $E$ evolves in the force field described by the function $V(x)$, something that the measuring instrument, here described as a cognitive entity, will have to ponder about every time it gives an unpredictable answer.  

This interrogative approach allows, for example, to explain why in a double-slit experiment the observed fringe structure can appear, in particular the more intense central fringe \citep{Aerts2009,Aertsetal2020}. At present, no similar analysis has been provided to explain in a conceptualistic way the main non-classical features of a quantum scattering process, for example the emergence of resonance phenomena. Here, however, our intention is different and does not depend on the details of the possible cognitive-like process of the measuring instruments, but on the fact that the outcomes result from such a process. From a cognitive point of view, the two possible responses are weighed together, in an overall evaluation process that is inherently holistic in nature. Therefore, it is natural to expect that the time needed to produce an answer does not directly depend, on average, on the answer given. Indeed, cognitively speaking, answering ``transmitted'' should not require more time than answering ``reflected,'' when the actualization is the result of a symmetry breaking (tension-reduction) process that incorporates all the possibilities taken into consideration \citep{asdb2014}. 

More generally, the time required to actualize an answer is generally not directly related to the semantic content of the answers. Consider the example of human cognition. If a group of people in Brussels are asked to choose whether to go to Rome or to Moscow the next day, we would not expect the time needed, on average, to answer ``Moscow'' to be longer than the time needed, on average, to select the answer ``Rome,'' just because Moscow is farther away in space from Rome than from Brussels. This is because the cognitive process of selecting one of the two options is not influenced by the spatial distances associated with them, but rather by the ``affective distances'' associated with the attractiveness of these places to the persons making the choice. 

Of course, what we propose here has general validity and goes beyond the specific context of a two-outcome quantum scattering process. It applies to any quantum measurement, provided that the experimental setup does not artificially favor the time of occurrence of certain outcomes/responses in some way, for example by placing a particular detector at a shorter distance from the measured system than the others. In effect, this would mean that the response represented by that detector would be seen before the others, causing the measuring cognitive entity to select it more quickly, on average, than the others. 

Similarly to what happens to us humans when we select an answer to a given question, in a situation where different answers are possible and compete with each other, a quantum measurement should also be understood as a ``vertical'' process in which a non-spatiotemporal conceptual entity passes from an abstract superposition state to a concrete eigenstate. Our classical prejudice pushes us, instead, to understand it as a ``horizontal'' process that takes place only within our spatiotemporal reality layer, characterized by maximally concrete states, hence the impossibility of making sense of experiments such as those that confirm the validity of the Hartman effect. 
 
Of course, reasonings of a similar nature can also apply outside of the conceptuality interpretation, for example by noting that when the energy of the incoming entity is well peaked, it can be viewed as a ``static'' spatially distributed phenomenon, potentially and jointly present in every region of space, hence time-delays would be mostly governed by the measuring procedure itself and the associated instantaneous collapse of the wave function; see \citep{Long2018} for an example of such an approach.

\section{Conclusions}
\label{Conclusions}

\noindent In this article we have re-analyzed the quantum tunneling effect and its temporal dynamics. Using the notion of conditional delay time, which maintains a consistent interpretation even for a quantum process, we have shown that there is a perfect equality between the time-delays for the reflected and transmitted outcomes, consequence of the unitarity of the scattering matrix. From this remarkable property it can be deduced that, as is the case for the reflected particle, also the transmitted one does not penetrate into the central region of the potential barrier. But since transmission remains a possible event, it also follows that transmission (in this case via tunneling) cannot be described as a process that takes place in space, or only in space. In other words, the tunneling process is a further confirmation that the nature of the microscopic quantum entities is non-spatial, and more generally non-spatiotemporal.

We also recalled that there is an interpretation of quantum mechanics, called the conceptuality interpretation, for which the notion of non-spatiality is central, as the entities of the micro-world are described in it as entities of a conceptual nature (but distinct from human concepts) and the measuring devices as cognitive-like entities sensitive to the meaning they carry. In such an explanatory scheme, non-spatiality becomes synonymous with abstractness, and spatiality becomes synonymous with (maximal) concreteness. In the conceptuality interpretation, the fact that the time-delays for transmission and reflection are identical is therefore something to be expected, since the process that leads to the actualization of an outcome can be understood as a cognitive process of an interrogative kind that produces a non-predetermined answer, and such process, of passage from an abstract state to a concrete state of the conceptual-like entity, would not take place within the  spatiotemporal layer, or primarily within the  spatiotemporal layer, but along a direction which, in a sense, is orthogonal to it.

\end{document}